\documentclass[aps,prb,preprint,groupedaddress,showpacs,amsmath,amssymb]{revtex4-1}
\usepackage{graphicx}
\usepackage{dcolumn}
\usepackage{bm}
\usepackage{float}

\begin{document}


\title{ Study of magneto-transport properties and quantum oscillations in PbSe single crystals }


\author{Naween Anand}
\affiliation{Department of Physics, University of Florida, Gainesville, FL 32611-8440, USA}
\author{C. Martin}
\affiliation{Department of Physics, Ramapo College, Mahwah, NJ 07430, USA}
\author{Genda Gu}
\affiliation{Condensed Matter Physics and Materials Science Department, Brookhaven National Laboratory, Upton, New York 11973-5000, USA}
\author{D. B. Tanner}
\affiliation{Department of Physics, University of Florida, Gainesville, FL 32611-8440, USA}


\date{\today}

\begin{abstract}
PbSe is a low-gap semiconductor with excellent infrared photo-detection properties. Here we report our high magnetic field and low temperature electrical properties measurement performed on a moderately doped PbSe single crystals with p-type bulk carrier density of around $1\times10^{18}$ cm$^{-3}$. Longitudinal resistance (R$_{xx}$) and Hall resistance (R$_{xy}$) were simultaneously measured between 0 T--18 T at several temperatures between 0.8 K--25 K. These transport measurements start showing oscillatory behavior around and above 6 T of magnetic field. The quantum oscillation frequency is ~15 T, giving an estimate for the carrier density of each L pocket in the BZ participating in these oscillations. The effective mass of the free carriers is estimated from the temperature dependence of oscillation amplitudes. Measurements as the magnetic fields is rotated reveal the magneto-transport properties of a 3D-like fermi surface. Free carrier scattering rate from the transport data has been estimated which also gave an estimation of Dingle temperature in PbSe. Room temperature optical measurements has been conducted and other optical properties has been estimated. The Drude-Lorentz model fit has been performed which show a low frequency phonon mode around 45 cm$^{-1}$ and bandgap of around 0.2 eV along with other interband electronic transitions. Some of the optical and transport results has been compared against other members of class IV-VI chalcogenides.
\end{abstract}
\maketitle

\section{INTRODUCTION}

PbSe is a narrow-gap semiconductor have been of great interest for many decades for their use in infrared optoelectronics and thermoelectric devices.\cite{Nimtz} PbSe has been found to have high dielectric constant and quite unusual infrared and electronic properties.\cite{Otfried} An upserge in interest  in  group  IV-VI  compounds has been stimulated because of the observation of a new topological class among these compounds coined as topological crystalline insulator. PbSnSe, PbSnTe and SnTe are some of the examples of TCIs. Previous studies of PbSe have reported the rock salt crystal structure at ambient temperature and pressure with a lattice parameter of $a = 6.13~${\AA} and a direct minimum energy band gap of around 0.28 eV at the L point in the Brillouin Zone.\cite{Delin,Otfried,Harman} It has L${}_{6}^{-}$ symmetry for the conduction band  while the valence band symmetry is denoted by L${}_{6}^{+}$, a topologically trivial phase.\cite{Dimmock}. PbSe is generally synthesized in doped state due to stoichiometric imbalance and due to high carrier mobilities at low temperatures in Pb salts, these carriers do provide an excellent way to study their dynamics, band transport parameters and Fermi surface geometry at low temperatures.

When a magnetic field is applied in a crystalline solid, electronic bands undergo through an additional quantization, known as the Landau quantization of the energy states. It modulates the density of states periodically as a function of the magnetic field. These modulations happen differently in different kind of electronic systems. For example, in ordinary metals these Landau states are evenly spaced with the energy gap proportional to the magnetic field and independent of the Fermi level. For Dirac fermions however, these Landau states are not evenly spaced. Energy gap becomes proportional to the square root of the magnetic field, depends on the Fermi level and Landau quantum number.\cite{shoenberg} In any case, these periodic modulations leads to oscillatory behaviour in several materials properties termed as quantum oscillations. In particular, this oscillatory behaviour in the longitudinal and the transverse conductivity is called Shubnikov-de Haas (SdH)oscillations. In this article, we report on SdH measurements performed primarily on a p-type PbSe single crystals.

\section{EXPERIMENTAL PROCEDURES}
Single crystals of PbSe were grown by a modified floating-zone method. High-purity (99.9999\%) elements of Pb and Se were loaded into a quartz ampoule and sealed under vacuum. The materials in the double-sealed quartz tube first were melted at $1000^{\circ}$~C in a box furnace and fully rocked to achieve homogeneous mixture. Then, the  premelt ingot rod (12 mm in diameter and 20 cm in length) located in a quartz tube was mounted in a floating-zone furnace. In the floating-zone furnace, the ingot was first premelted at a velocity of 200 mm/hr and then grown at 1.0 mm/hr in a 1 bar Ar atmosphere. The crystals had the rock salt structure with lattice parameter $a$ = 6.13~{\AA} as shown in the figure 1. These crystals are opaque and have a metallic luster. They are brittle and easily cleave along the (100) plane. A single crystal of size $4\times3\times0.4$ mm${}^{3}$ with smooth (100) crystal plane as the exposed surface was selected for transport measurements.
\begin{figure}[H]
\centering
\includegraphics[width=5 in,height=5 in,keepaspectratio]{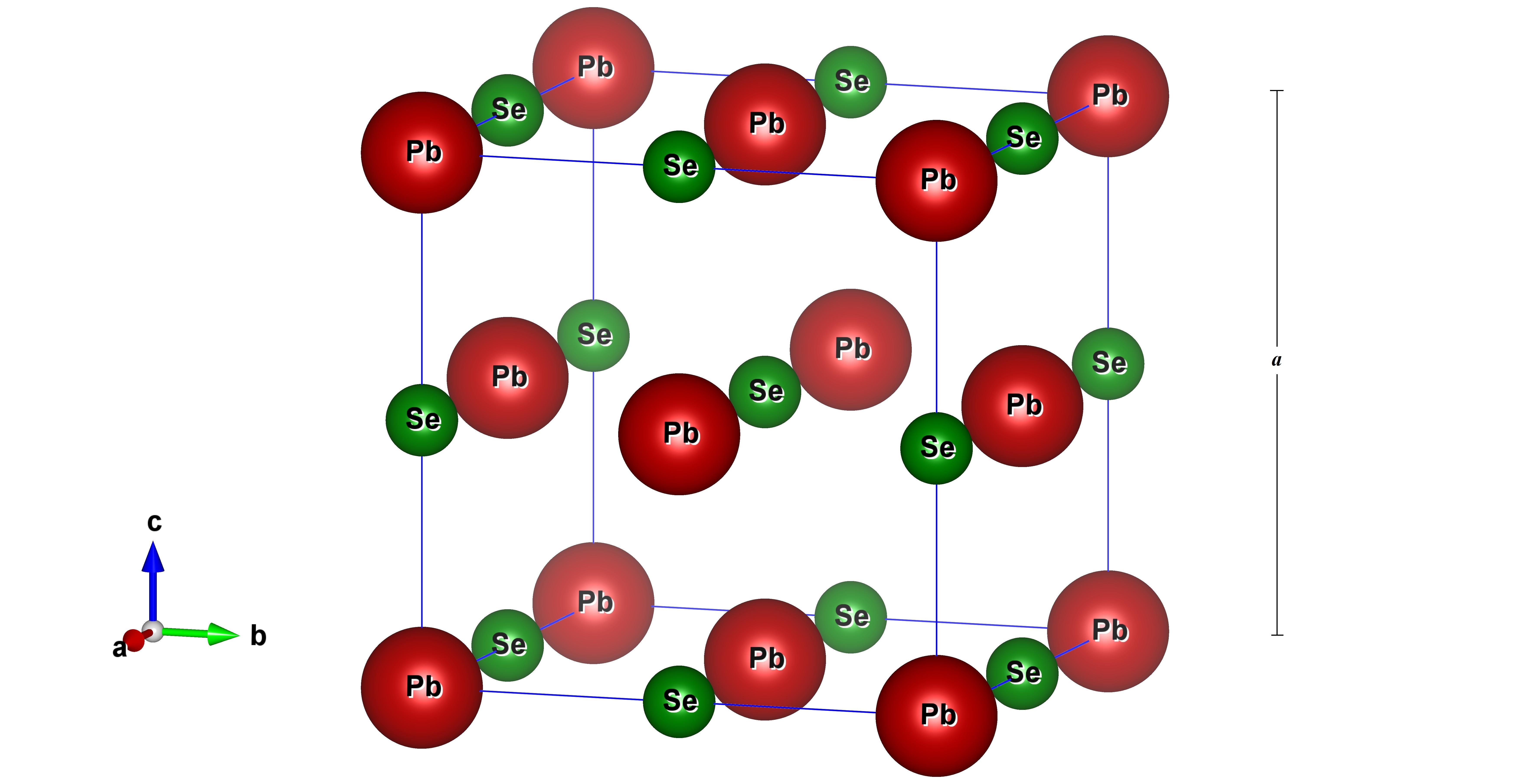}
\caption{\label{fig:Crystalstructurepbse} (Color online) Rock-salt crystal structure of a PbSe single crystal.}
\end{figure}
Low temperature and high magnetic field electrical properties measurements were performed in the SCM-2 facility at the National High Magnetic Field (NHMFL) in Tallahassee.
\begin{figure}[H]
\centering
\includegraphics[width=5 in,height=5 in,keepaspectratio]{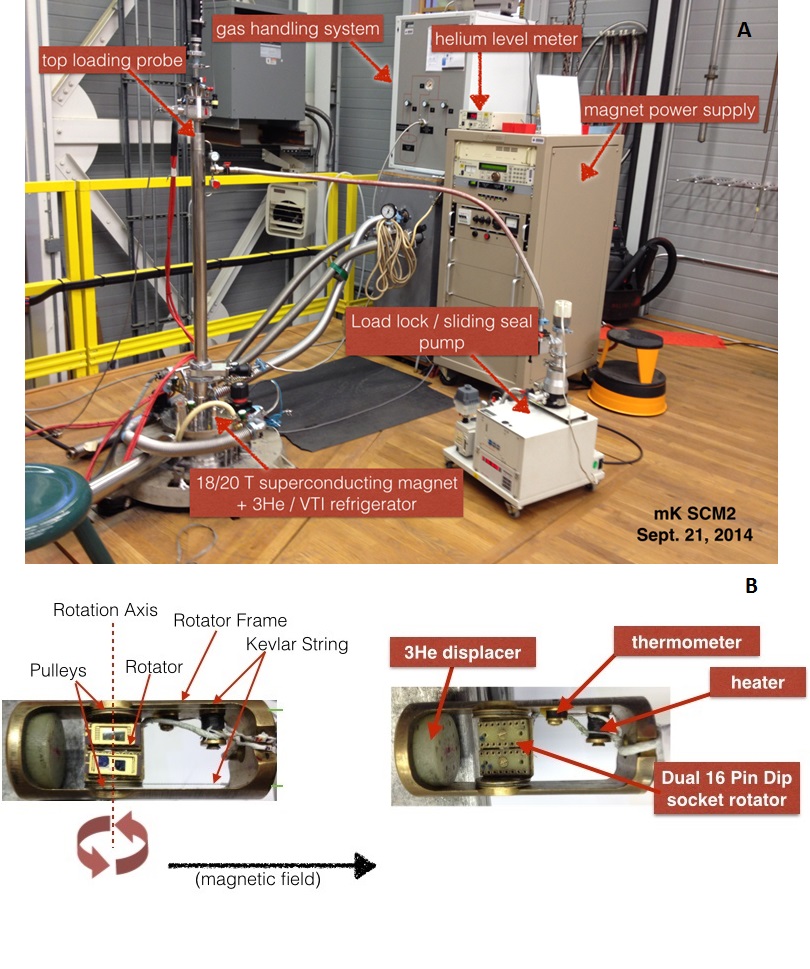}
\caption{\label{fig:NHMFL} (Color online)  A) 18/20 Tesla General Purpose Superconducting Magnet. B) Rotating probe piece in conjunction with the sample end of Cryostat. (Pictures taken from the NHMFL website under subsection  \textsc{\char13}\textsc{\char13}SCM2 He-3 System \& VTI\textsc{\char13}\textsc{\char13})}
\end{figure}
The facility consists of a top loading He-3 cryostat, with sample in liquid and a base temperature of 0.3 K, and an  18/20  Tesla  YBCO based superconducting  magnet. Magnetic fields up to 18~T were applied perpendicular to the (100) plane while the Hall resistance transverse to input current and magnetic field and longitudinal resistance along the input current direction were simultaneously measured. Gold wires were attached using silver paint and the sample resistance was measured using a commercial resistance bridge. The sample temperature were varied between 0.85 K and 25 K in order to measure the change in the electrical properties. The Hall-effect measurements gave the sign (p-type) and value for the carrier density $n$ in this temperature range. Errors in the carrier density were estimated from the non-uniform thickness of the sample, non-rectangular cross section, and contact size.  Samples  were mounted on a rotating probe with an angular resolution better than 1$^{\circ}$. After temperature dependent measurements, the probe was rotated so that the normal direction to the (001) plane makes angle varying from zero to 90$^{\circ}$  with the magnetic field. Electrical measurements at these angles were measured at the base temperature of 0.85 K.The experimental set-up of magnet system and rotating probe piece illustrating the interfacing of cryostat with sample holder, electronics and rotating piece is shown in fig. 2.

On the other hand, Fig. 3(A) shows the PbSe sample installed on 16-pin dip with 6 solid contacts using gold wire and silver paint. In the schematic diagram on the right of Fig. 3(B), contacts $a$ and $b$ are the current input and output ends. Contacts $c$ and $d$ together are used for longitudinal resistance measurement whereas contacts $e$ and $f$ together were used for Hall resistance measurement. Next figure 4 shows the interfacing schematics between the cryogenic system, the relevant electronics for the temperature and electric signal detection and control and also the computer-CPU for controlling the experiment.

Soon after the transport measurement, room temperature reflectance measurement were conducted using a Bruker 113v Fourier-transform interferometer at University of Florida. A helium-cooled silicon bolo\-meter detector was used in the 40--650 cm${}^{-1}$ spectral range and a DTGS detector was used
from 600--7000 cm${}^{-1}$. Higher frequency response up to 35,000 cm$^{-1}$ was measured using a Zeiss microscope photometer. Because the rock salt structure of the material implies isotropic optical properties, all optical measurements were performed using non-polarized light at near-normal incidence on the (100) crystal plane.

\begin{figure}[H]
\centering
\includegraphics[width=4 in,height=4 in,keepaspectratio]{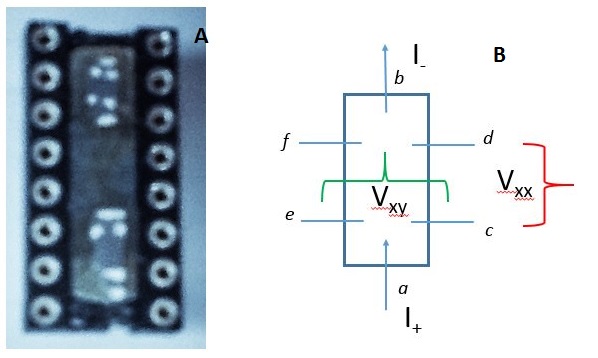}
\caption{\label{fig:contacts} (Color online)  A) Actual sample mounted on a 16 pin dip with 6 separate contacts. B) Schematics for the longitudinal and Hall resistance measurement.}
\end{figure}
\begin{figure}[H]
\centering
\includegraphics[width=3.4 in,height=3.4 in,keepaspectratio]{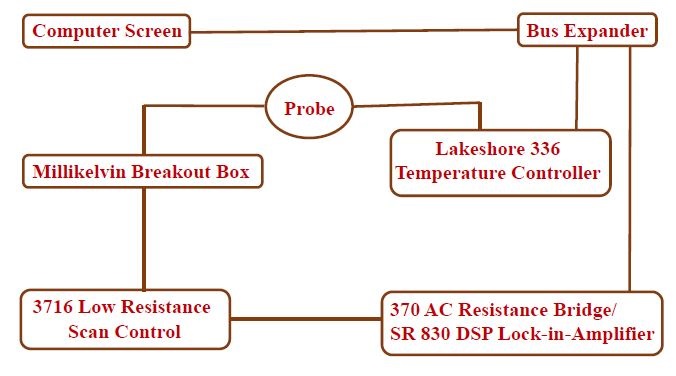}
\caption{\label{fig:setup} (Color online) Interfacing schematics for signal detection.}
\end{figure}
\section{EXPERIMENTAL RESULTS AND ANALYSIS}
\subsection{Transport Study}
Temperature dependent high magnetic field longitudinal resistance and Hall resistance is shown in figure 5 and figure 6 respectively. Apart from large positive magnetoresistive trend, we also observe some non-linear modulation in both graphs above 6 T of magnetic field. This modulation is very obvious at base temperature of 0.85 K and feature weakens as temperature increases.
\begin{figure}[H]
\centering
\includegraphics[width=3.4 in,height=3.4 in,keepaspectratio]{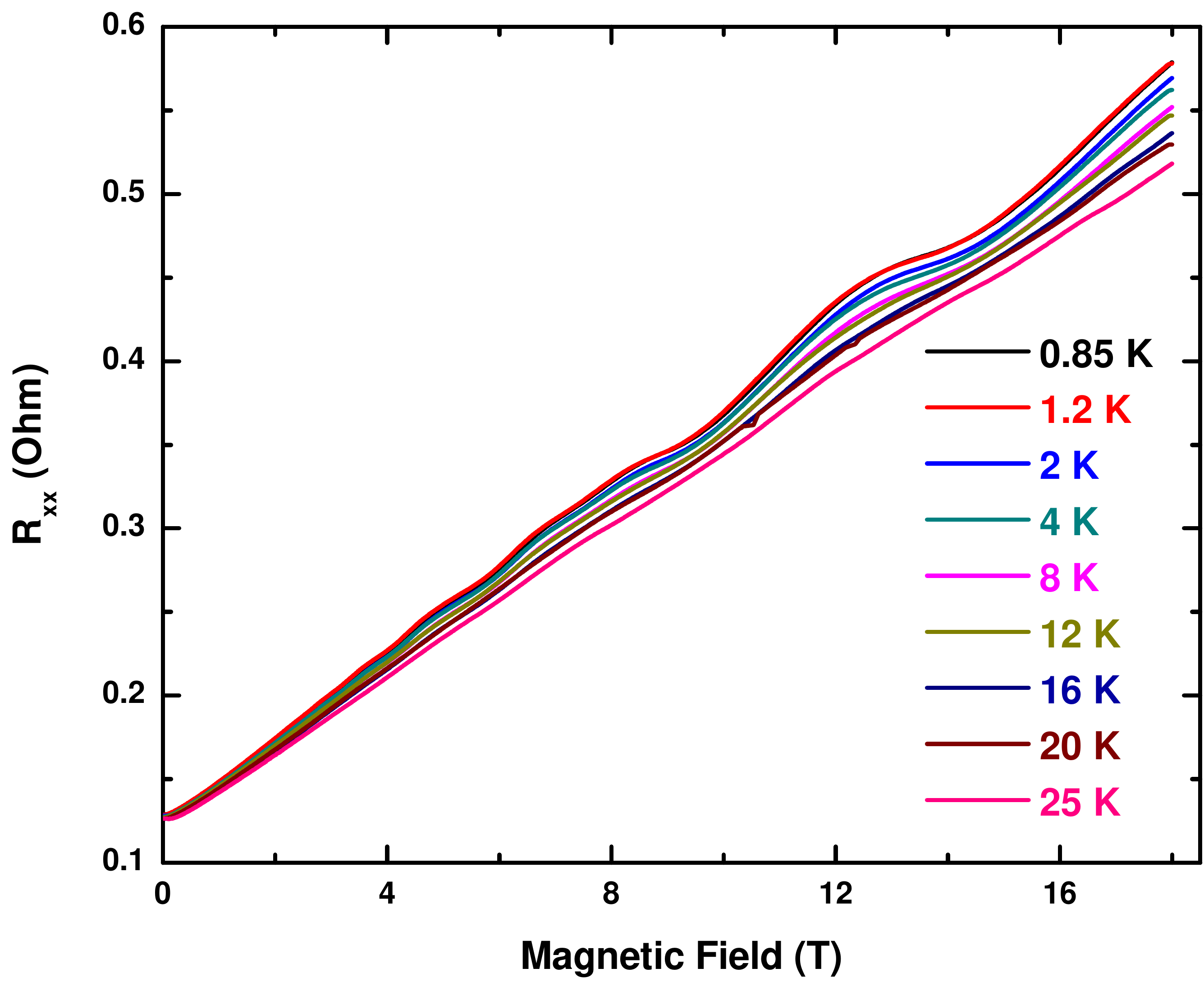}
\caption{\label{fig:LO} (Color online) Temperature dependent longitudinal resistance as a function of magnetic field.}
\end{figure}
\begin{figure}[H]
\centering
\includegraphics[width=3.4 in,height=3.4 in,keepaspectratio]{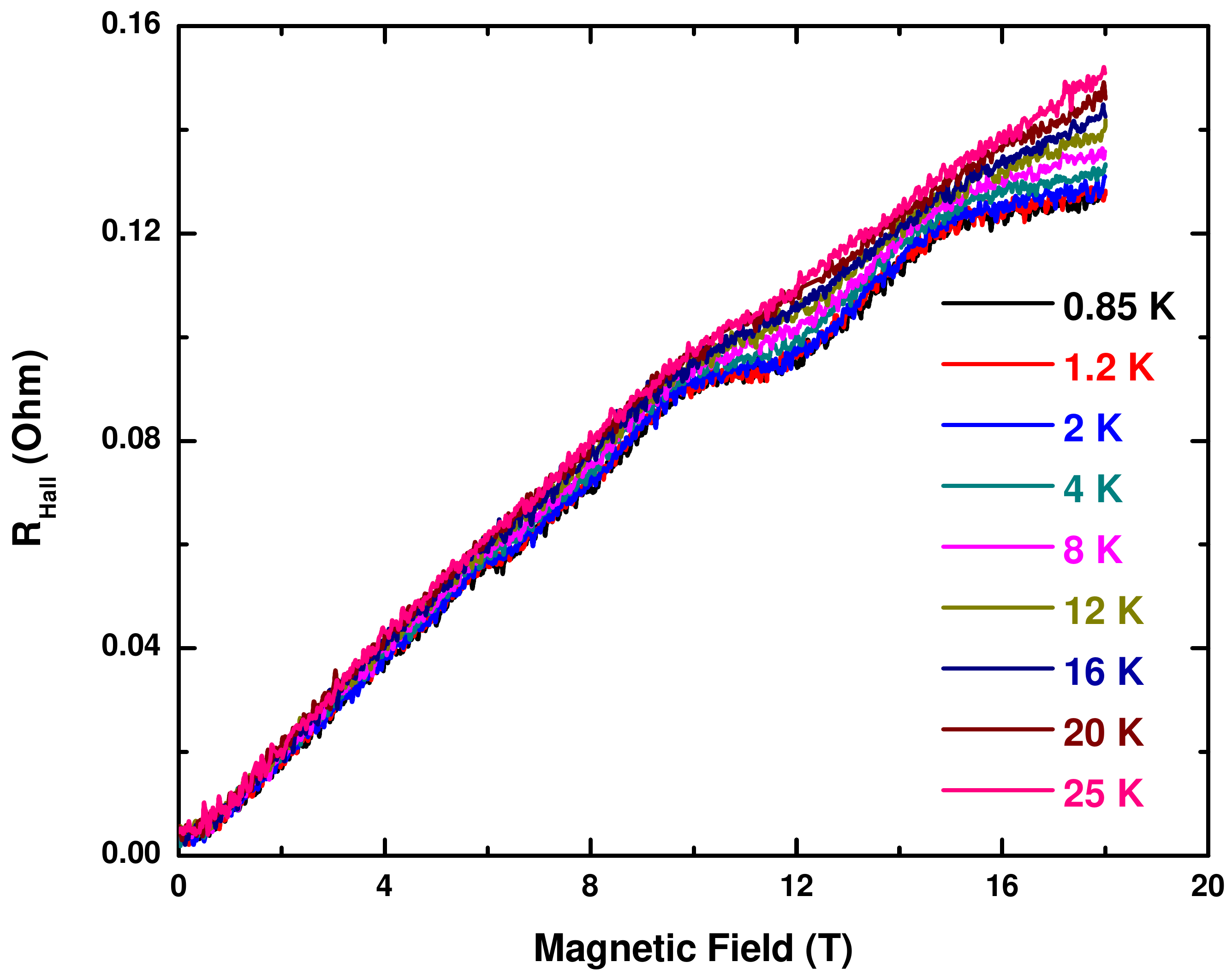}
\caption{\label{fig:TO} (Color online) Temperature dependent Hall resistance as a function of magnetic field.}
\end{figure}
After subtracting a second order polynomial background fit from R$_{xx}$ and R$_{Hall}$, oscillation amplitudes are plotted against inverse field (1/B). This oscillatory modulation in resistance amplitudes, periodic in 1/B can be directly observed in figure 7 and 8. Signal to noise ratio for longitudinal oscillation data is much better as compared to Hall oscillation data which explains much clearly resolved oscillation periodicity in 1/B. Nonetheless, oscillatory behavior is quite clear in both graphs. Few obvious trends could be seen in these figures. Firstly, oscillation amplitude appears to be periodic in 1/B. Secondly oscillation amplitude decreases as inverse field increases and lastly, oscillation amplitude decreases as temperature increases without any shift in the crests and troughs position.

Hall resistance data is smoothed using a Savitzky-Golay filter\cite{Savitzky,Manfred} as shown in figure 9 in order to enhance the oscillatory feature buried within random noise which essentially performs a local polynomial regression to determine the smoothed value for each data point while preserving features of the data such as peak height and width. As compared to simply averaging points in which features gets washed away, this is a much better procedure which perform a least squares fit of a small set of consecutive data points to a polynomial and take the calculated central point of the fitted polynomial curve as the new smoothed data point.
\begin{figure}[H]
\centering
\includegraphics[width=4.5 in,height=4.5 in,keepaspectratio]{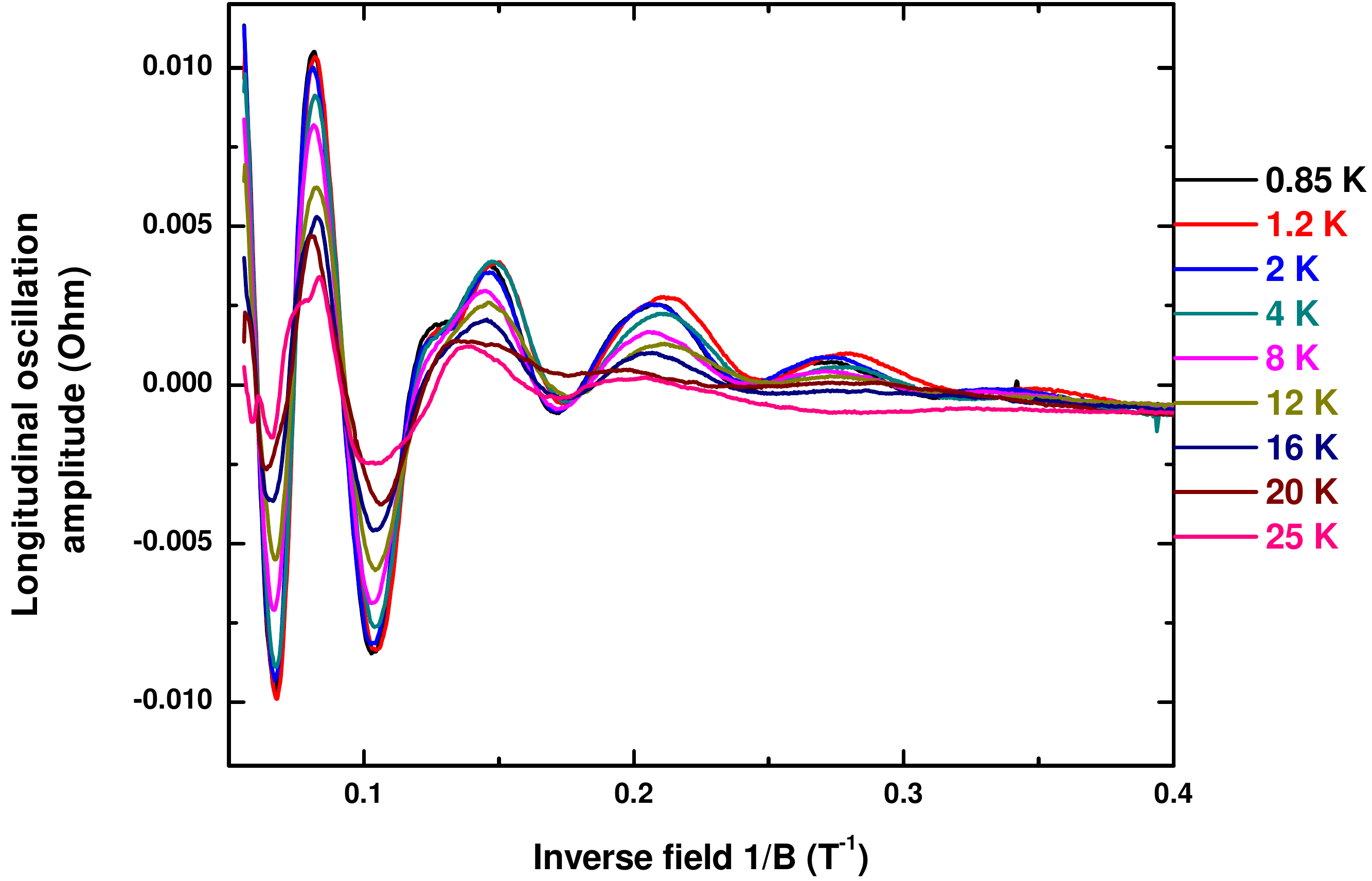}
\caption{\label{fig:LOA} (Color online) Temperature dependent longitudinal resistance oscillation as a function of inverse magnetic field.}
\end{figure}
\begin{figure}[H]
\centering
\includegraphics[width=3.4 in,height=3.4 in,keepaspectratio]{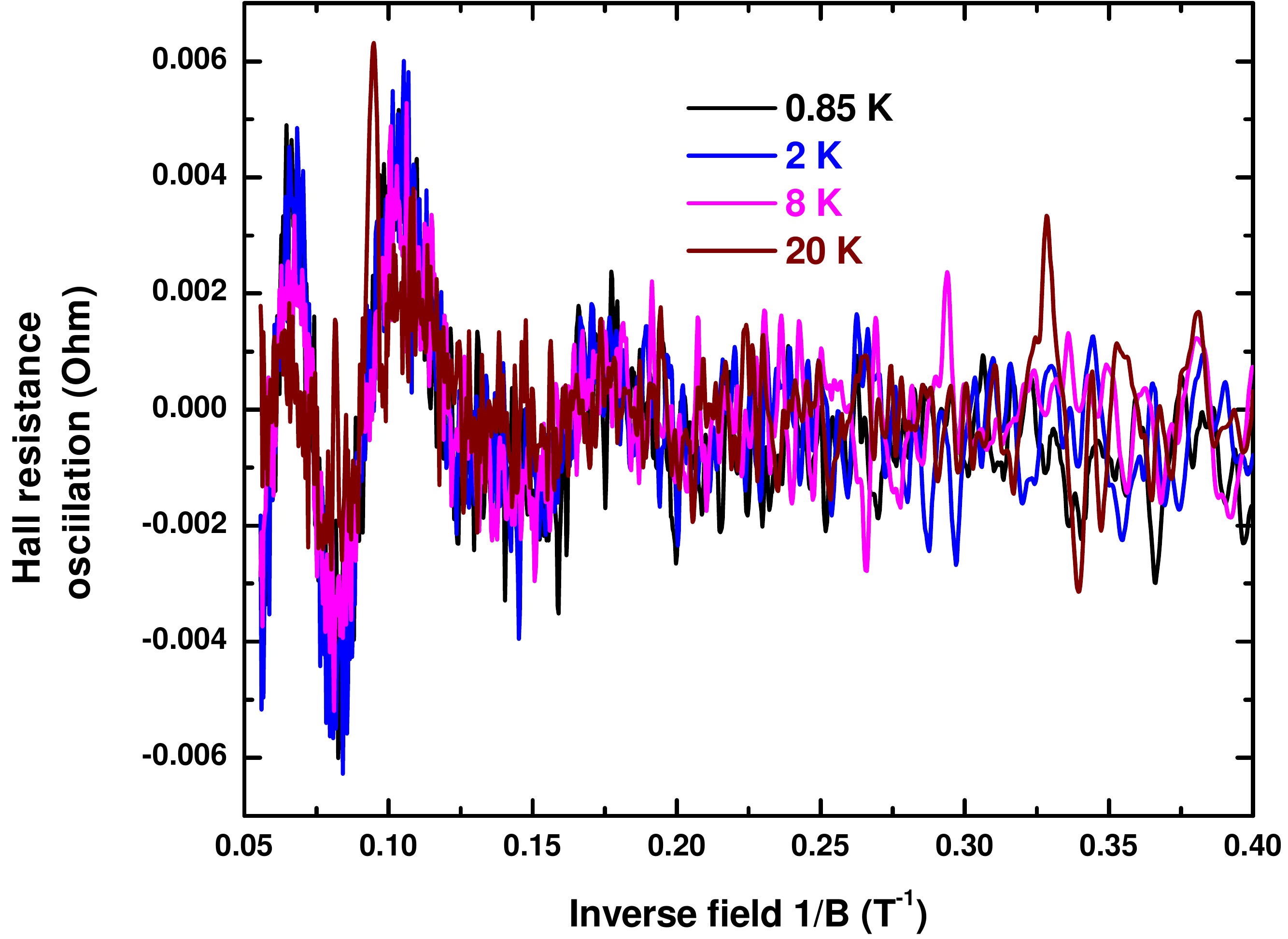}
\caption{\label{fig:TOAE} (Color online) Temperature dependent Hall resistance oscillation as a function of inverse magnetic field before smoothening.}
\end{figure}
\begin{figure}[H]
\centering
\includegraphics[width=4.5 in,height=4.5 in,keepaspectratio]{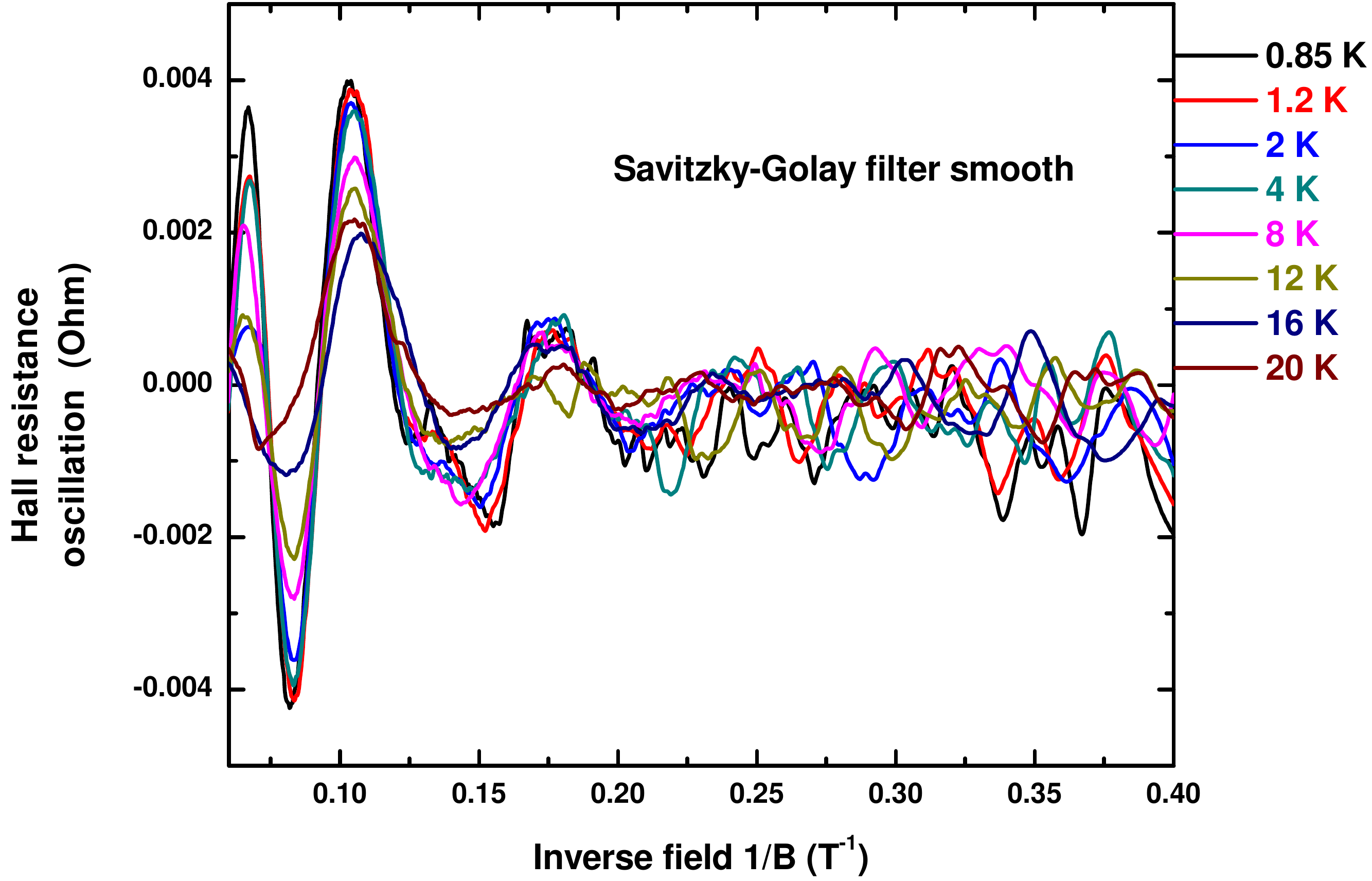}
\caption{\label{fig:TOAS} (Color online) Temperature dependent Hall resistance oscillation as a function of inverse magnetic field after smoothening.}
\end{figure}
Fast Fourier transform (FFT) of longitudinal and Hall data is shown in figure 10. Both plots indicate a highest peak of frequency of oscillations  at the frequency of 14.7 T for the field applied normal to the sample surface. FFT plot of Hall data is noisy because of the poor signal to noise ratio of the original Hall data.  A secondary peak is also observed at around 29.5 T which corresponds to the second harmonic of the fundamental peak. This higher harmonic may be present there because of the Zeeman-splitting of each landau level at high magnetic field. This splitting increases the oscillation frequency by almost 2-fold. FFT plots at different temperatures show no shift in the peak position meaning frequency of oscillation is independent in this temperature range.

Figure 11 gives information about the phase relationship between longitudinal oscillation and Hall oscillation at the base temperature of 0.85 K. Measured Hall data has poor signal to noise ratio and hence is smoothed using Savitzky-Golay filter and Fourier transform filter and then compared with Longitudinal oscillation. The phase of the oscillations is such that the decrease (increase) in the longitudinal resistance is accompanied by an increase (decrease) in the absolute value of the Hall resistance. This trend is quite evident at high magnetic field when oscillation amplitude is large enough and both signals are almost 180$^{\circ}$ out of phase. As oscillation amplitude decreases with decreasing magnetic field phase relation becomes ambiguous. This could be due to the poor signal to noise ratio of the Hall oscillation data.
\begin{figure}[H]
\centering
\includegraphics[width=4.5 in,height=4.5 in,keepaspectratio]{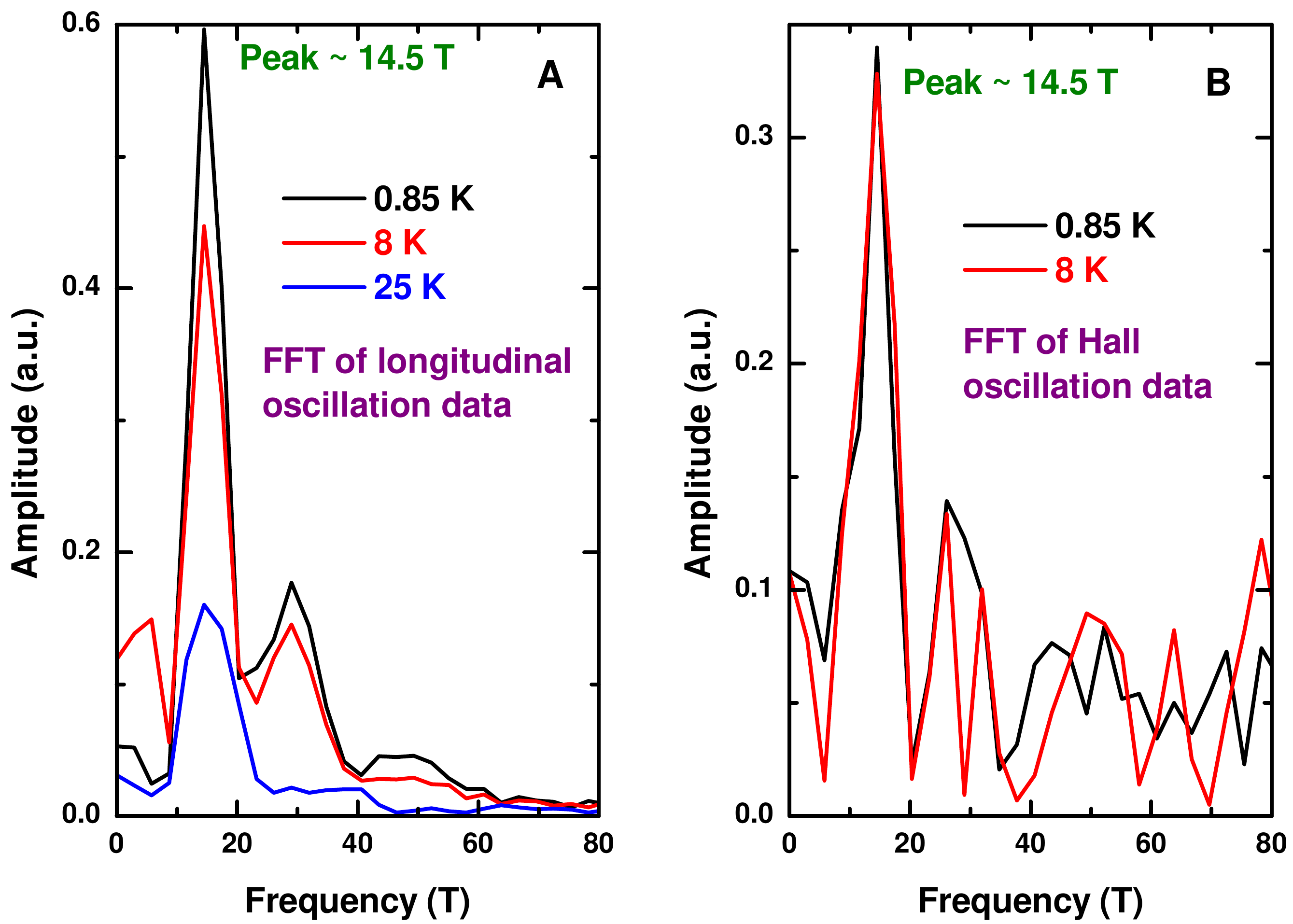}
\caption{\label{fig:fft} (Color online) A) Temperature dependent FFT of longitudinal resistance oscillation data. B) Temperature dependent FFT of Hall resistance oscillation data.}
\end{figure}
\begin{figure}[H]
\centering
\includegraphics[width=3.4 in,height=3.4 in,keepaspectratio]{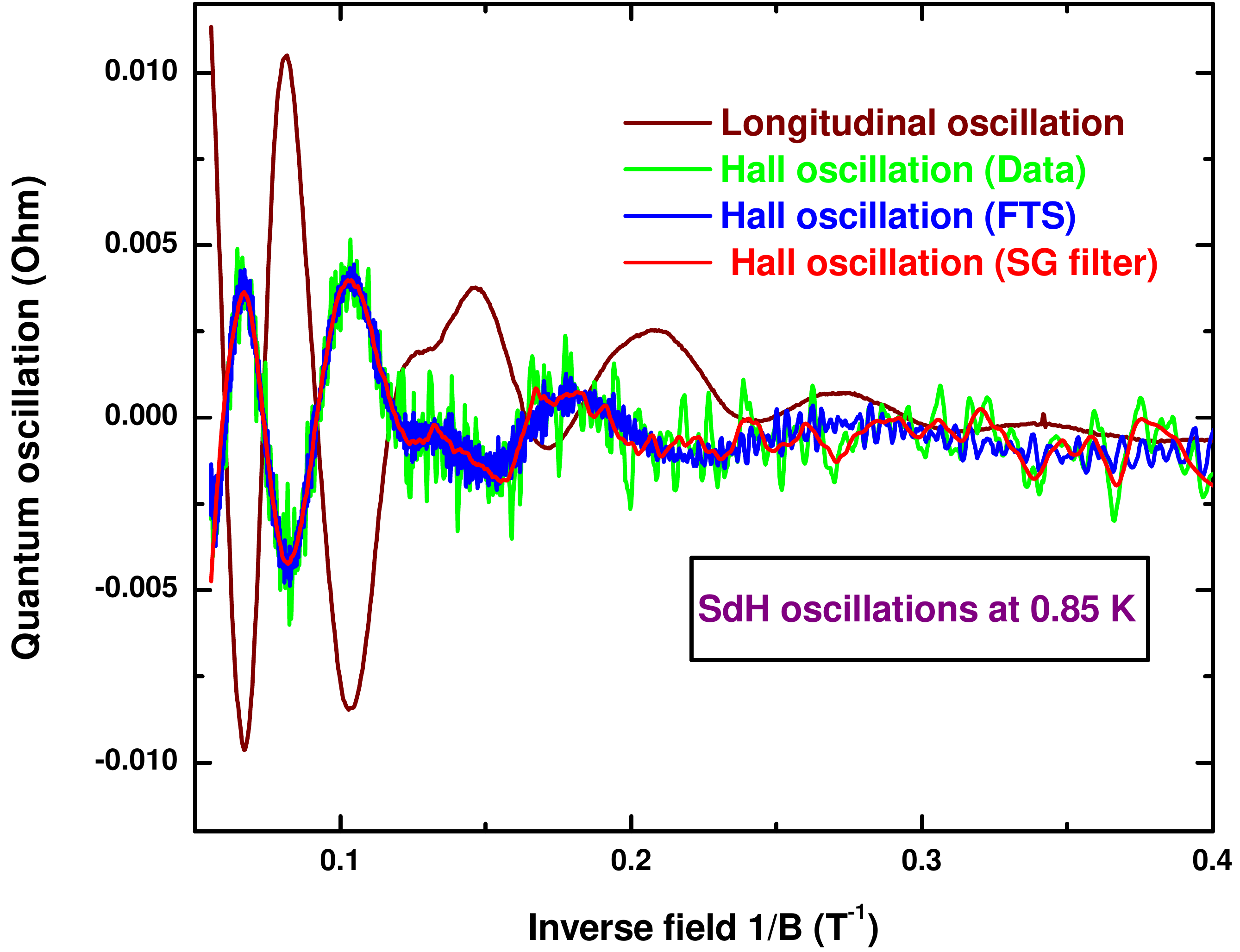}
\caption{\label{fig:phase} (Color online) Phase relation between longitudinal and Hall quantum oscillations at 0.85 K.}
\end{figure}

The exact expression for $\rho$$_{xx}$ and $\rho$$_{xy}$, derived by inverting the ($2\times2$) $\sigma$ tensor is given as

\begin{subequations}
\begin{align}
{} & \rho_{xx}=\frac {\sigma_{xx} }{\sigma_{xx}^{2}+\sigma_{xy}^{2}}\\
& \rho_{xy}=\frac {\sigma_{xy} }{\sigma_{xx}^{2}+\sigma_{xy}^{2}}
\end{align}
\end{subequations}

Knowing the dependence, a small change in $\sigma_{xx}$ and $\sigma_{xy}$ will produce changes in $\rho_{xx}$ and $\rho_{xy}$. The ratio of the change in resistivity components could be derived as

\begin{equation}
\frac {\Delta \rho_{xy}}{\Delta \rho_{xx}}=\frac {(2K)\Delta \sigma _{xx}-(K^{2}-1)\Delta \sigma _{xy}}{(K^{2}-1)\Delta \sigma _{xx}+(2K)\Delta \sigma _{xy}}
\end{equation}

 Here $K$ represents $\rho_{xx}/\rho_{xy}$. Under the actual experimental condition, the value of $K$ decreases as field increases. The value of K at and above moderate value of field seems to saturate between 5 and 4. This equation involving $K$ and the rate of change of conductivity components with changing field dictates whether the ratio is positive or negative. One can infer that at least at high magnetic field $\Delta\sigma_{xy}$ term dominates making the ratio negative and oscillations are 180$^{\circ}$ out of phase. At low field however, poor signal to noise ratio seems to dilute the phase correlation. This out of phase oscillation behavior is also reported in previous study of microwave radiation induced magneto-oscillations in the longitudinal and Hall resistance of a 2-D electron gas.\cite{Studenikin} However a theoretical determination of the phase relationship between  $\Delta\rho_{xx}$ and $\Delta\rho_{xy}$ has not yet been addressed directly.\cite{Durst}

SdH oscillations provide a means to selectively and quantitatively characterize the 2-D surface states occupied by Dirac fermions that coexist with 3-D bulk states occupied by Schrodinger type non-Dirac fermions. These quantum oscillations are expressed in the Lifshitz-Kosevich theory as\cite{shoenberg}

\begin{equation}
\Delta \sigma_{xx}=A_{0}R_{T}R_{D}R_{S}\cos[2\pi(\frac F B -\frac 1 2+\beta )]
\end{equation}

Here, A$_{0}$ is a constant and other three coefficients R$_{T}$, R$_{D}$ and R$_{S}$ are called temperature, Dingle and spin damping factors respectively. F is the frequency of oscillation and $\beta$ is called the phase factor. The argument of the cosine function in the equation determines the local extrema positions of SdH oscillations plotted against 1/B. Local extrema points lie on these oscillation curves whenever

\begin{equation}
2\pi(\frac {F} {B_{N}} -\frac 1 2+\beta )=(2N-1)\pi
\end{equation}

Here N denotes the Landau level quantum number. This equation implies that if LL quantum number N is plotted against  1/B$_{N}$ on Y-axis, it should make a straight line with a slope 1/F where F corresponds to the oscillation frequency and intercept on the N-index axis gives the phase factor $\beta$. Figure 12shows the linear fit between N-index and 1/B$_{N}$. This analysis is also known as Landau-level fan diagram.

\begin{figure}[H]
\centering
\includegraphics[width=3.4 in,height=3.4 in,keepaspectratio]{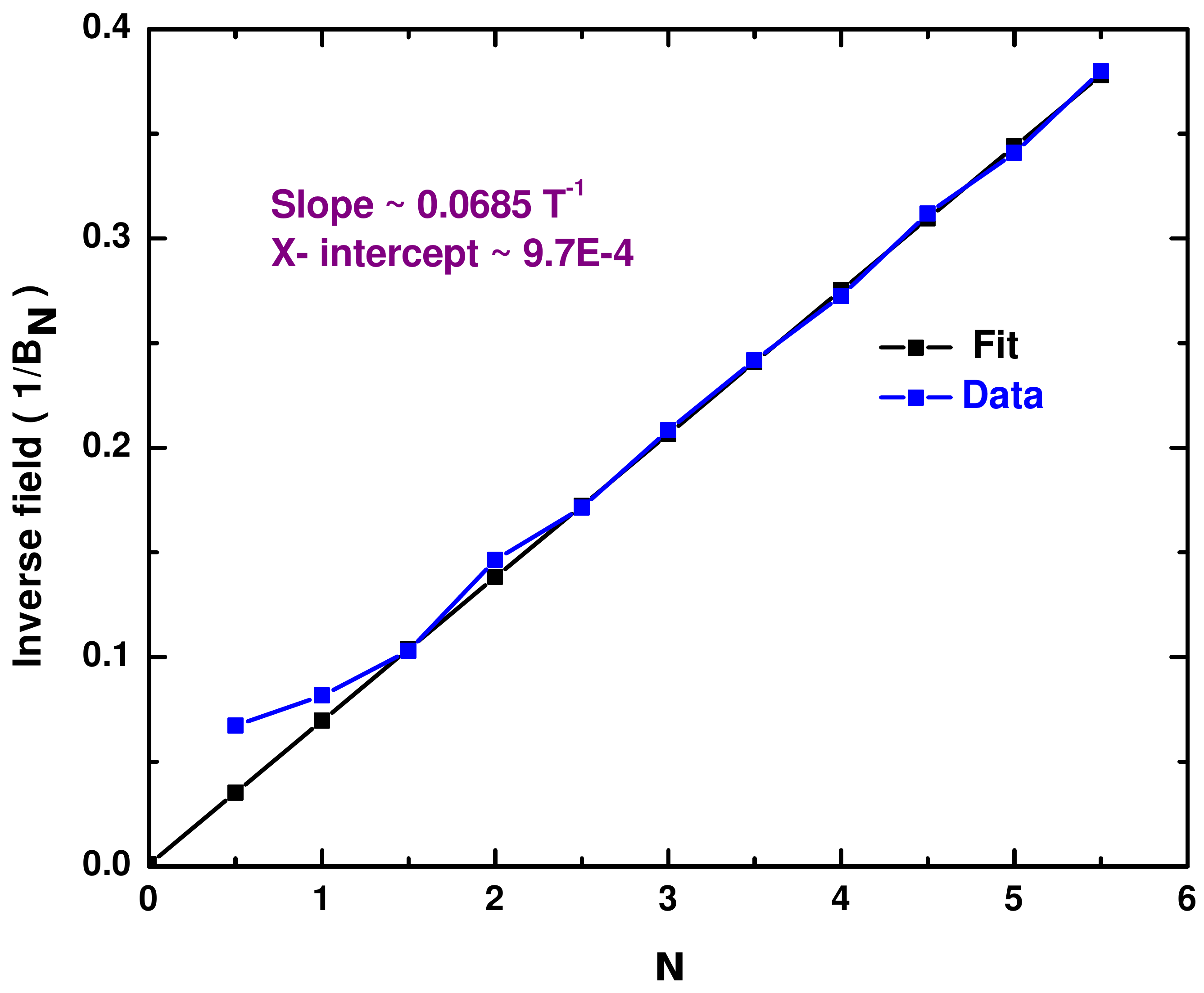}
\caption{\label{fig:LLFD} (Color online) Landau level fan diagram from longitudinal resistance oscillation data.}
\end{figure}

This parallel analysis of SdH oscillation data predicts an oscillation frequency of 14.6 T, very close to the FFT predicted oscillation frequency. Besides X-intercept which is close to zero (or integer) suggests that carriers engaged in oscillation phenomena is non-Dirac type bulk fermions. Assuming a circular cross section of FS, the frequency of oscillation F is related to the Fermi wave vector k$_{F}$ as

\begin{equation}
F=(\frac {\hbar} {2\pi e})\pi k_F^2
\end{equation}

As SdH oscillations come from a 3D bulk carriers and assuming that fermi surface is spherical, the carrier density could be expressed as

\begin{equation}
n_{SdH}=[\frac {2}{(2\pi)^3}]\frac {4\pi} {3}k_F^3=(\frac 1 {3\pi^2})(\frac {2eF}{\hbar})^{3/2}
\end{equation}

Using previous equation, the carrier density $n_{SdH}$ is estimated to be around $3.4\times10^{17} cm^{-3}$. Now considering the 4-fold degeneracy of the L-valley in PbSe, estimated bulk carrier density $n_{3D}$ should be compared with $4\times n_{SdH}$. Hall resistance measurement is another way of estimating bulk carrier density. The exact expression is given as

\begin{equation}
R_{Hall}=\frac {V_{Hall}}{I}=\frac {B}{n_{3D}et}
\end{equation}

Here $t$ is the thickness of the sample. Looking back figure 6, one can find the slope of the linear region between 0 T and 6 T range. Figure 13 gives the estimated bulk carrier density as a function of temperature.

\begin{figure}[H]
\centering
\includegraphics[width=3.4 in,height=3.4 in,keepaspectratio]{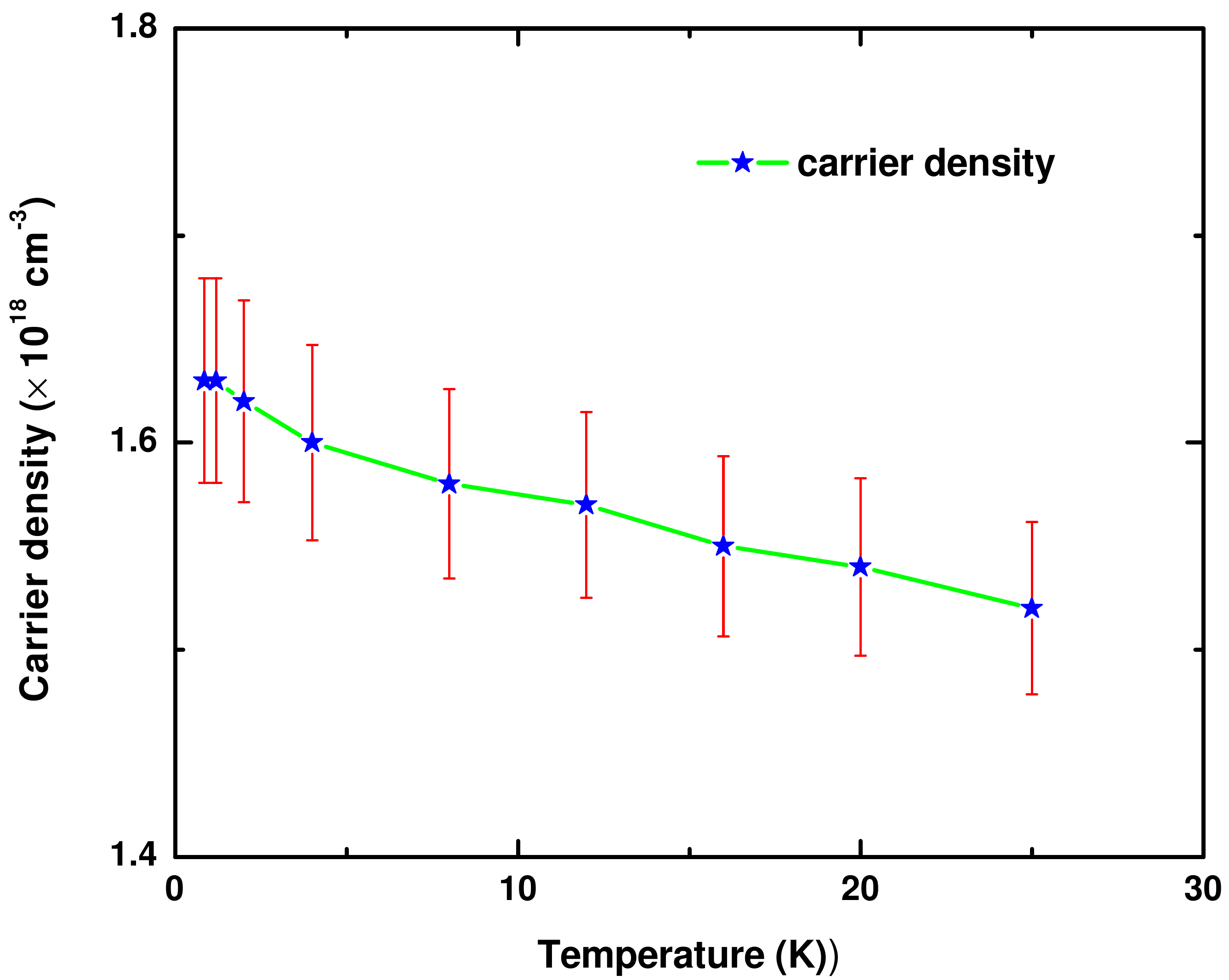}
\caption{\label{fig:n-T} (Color online) Temperature dependence of bulk carrier density derived from Hall resistance oscillation data.}
\end{figure}

Taking into account of 4 fold degeneracy of L-valley, $n_{3D}$ looks in good agreement with $n_{SdH}$. To  investigate  further  the  carrier  properties,  we  analyze   the  temperature dependence of the longitudinal oscillation amplitude of Landau level index N=1. Oscillations are visible up to at least 25 K and gets significantly damped due to the thermal broadening of the quantized Landau levels which is also evident if we look at the amplitude of the Fourier transform shown in figure 10. The oscillation amplitude is expected to follow the Lifshitz-Kosevich temperature factor dependence given as

\begin{equation}
\begin{aligned}
R_T={} & [\frac {\gamma T}{\sinh(\gamma T)}] \\
\gamma = & (\frac{14.69m^*}{m_{e}B})
\end{aligned}
\end{equation}

Here $B$ is the magnetic field, $m^{*}$ is the effective mass, and
$m_{e}$ is the rest mass of the electron.\cite{shoenberg} Figure 14 shows
that result of the fit to the above expression for the amplitude at an average magnetic field of $B \approx 11$~T.

\begin{figure}[H]
\centering
\includegraphics[width=3.4 in,height=3.4 in,keepaspectratio]{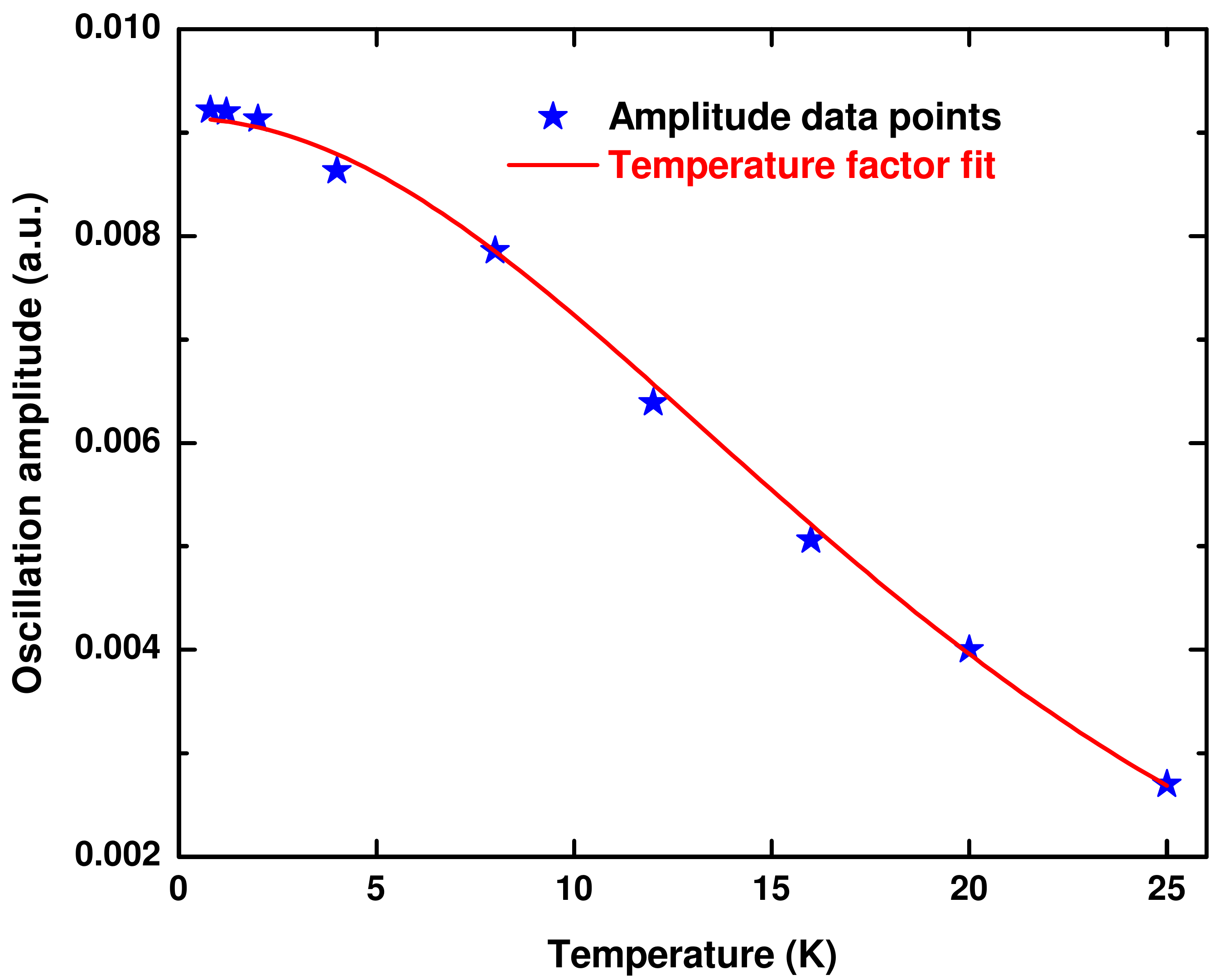}
\caption{\label{fig:m} (Color online) Amplitude of the oscillations at $B$ $\approx 11$~T for different temperatures and a fit to the temperature-dependent damping term as explained in the text.}
\end{figure}

The same analysis is repeated for different peaks of the SdH oscillations from longitudinal and the Hall resistance. Average carrier effective mass $m^{*}$ thus obtained is around $0.09m_{e}$. Knowing the carrier density and effective carrier mass, transport data allows us to derive transport scattering rate $1/\tau$. Figure 15 shows the scattering rate in the temperature range of 0.8 K--25 K.
Within the error bars, it is safer to report the thermal average of scattering rate $\left \langle 1/\tau \right \rangle_{T}$, came out around 26 cm$^{-1}$. This value looks reasonable if we compare the scattering rate of PbSnSe system discussed in the previous chapter. Based on the average transport scattering rate, Dingle temperature T$_{D}$=$\hbar/2\pi\tau k_{B}$ is estimated to be around 7 K. The corresponding average hole mobility $\mu = e\tau/m_h^*$ is around 3990 cm${}^2$/Vs.  Based on the average hole mobility value, the average optical conductivity $\sigma = n e \mu$ is estimated to be about 1020 $\Omega^{-1} cm^{-1}$ in the temperature range of 0.8 K--25 K.
\begin{figure}[H]
\centering
\includegraphics[width=3.4 in,height=3.4 in,keepaspectratio]{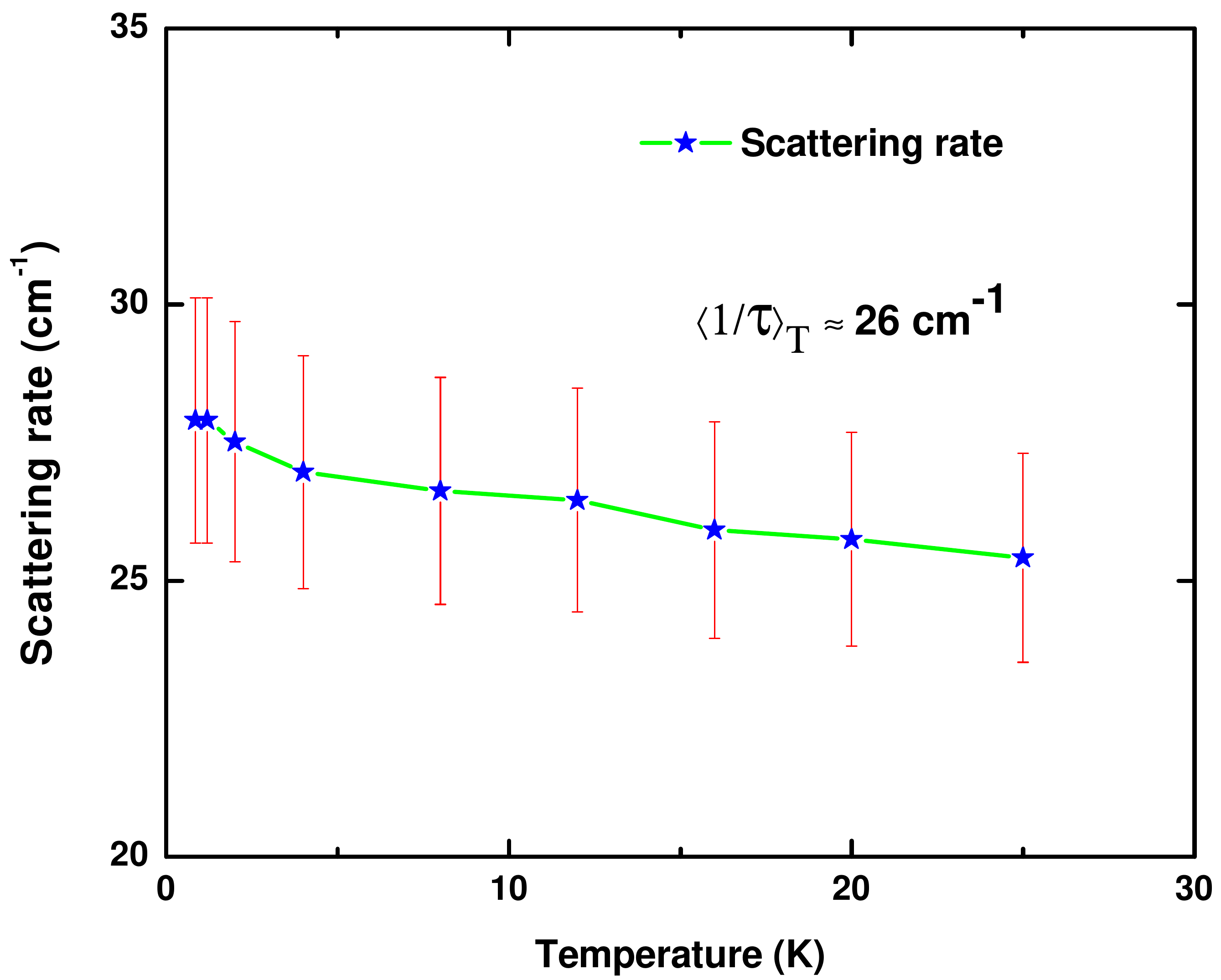}
\caption{\label{fig:scattering} (Color online) Carrier scattering rate as a function of temperature.}
\end{figure}

The angular dependence of the SdH oscillations provides a valuable insight into the dimensionality of the Fermi surface. In principle, for a 3D Fermi surface, electron orbits will be closed for any orientation of the magnetic field, and thus oscillations should be observed for any angle between the magnetic field and sample surface. However for a quasi-2D Fermi surface (e.g., a cylinder), as long as the oscillation frequency lies within the highest available magnetic field, would show oscillations up to relatively large angles. In contrast, a strictly 2D layer backed by a conducting bulk may loose orbital coherence at small angles if a tilted field drives carriers into the bulk. Figure 16 shows the longitudinal resistance for different tilt angle between the normal to the surface and the magnetic field.
After subtracting a second order polynomial background fit from R$_{xx}$, oscillation amplitudes are plotted against inverse field (1/B) in figure 17.
Fast Fourier transform (FFT) of longitudinal data is shown in figure 18 which gives oscillation frequency of about 14.5 T for angles less than 30$^{\circ}$. The oscillation frequency decreases to about 2.9 T for angles between 30$^{\circ}$ and 90$^{\circ}$. This change could also be realized in figure 16 when trend of oscillation changes around and above 30$^{\circ}$. For a 2-D Fermi surface, oscillation should ideally vanish. Since oscillation sustains for higher angles reveals the 3-D nature of the Fermi surface. Highly 3-D Fermi surface of p-type PbSe is also predicted in previous study.\cite{Svane}

\begin{figure}[H]
\centering
\includegraphics[width=4.5 in,height=4.5 in,keepaspectratio]{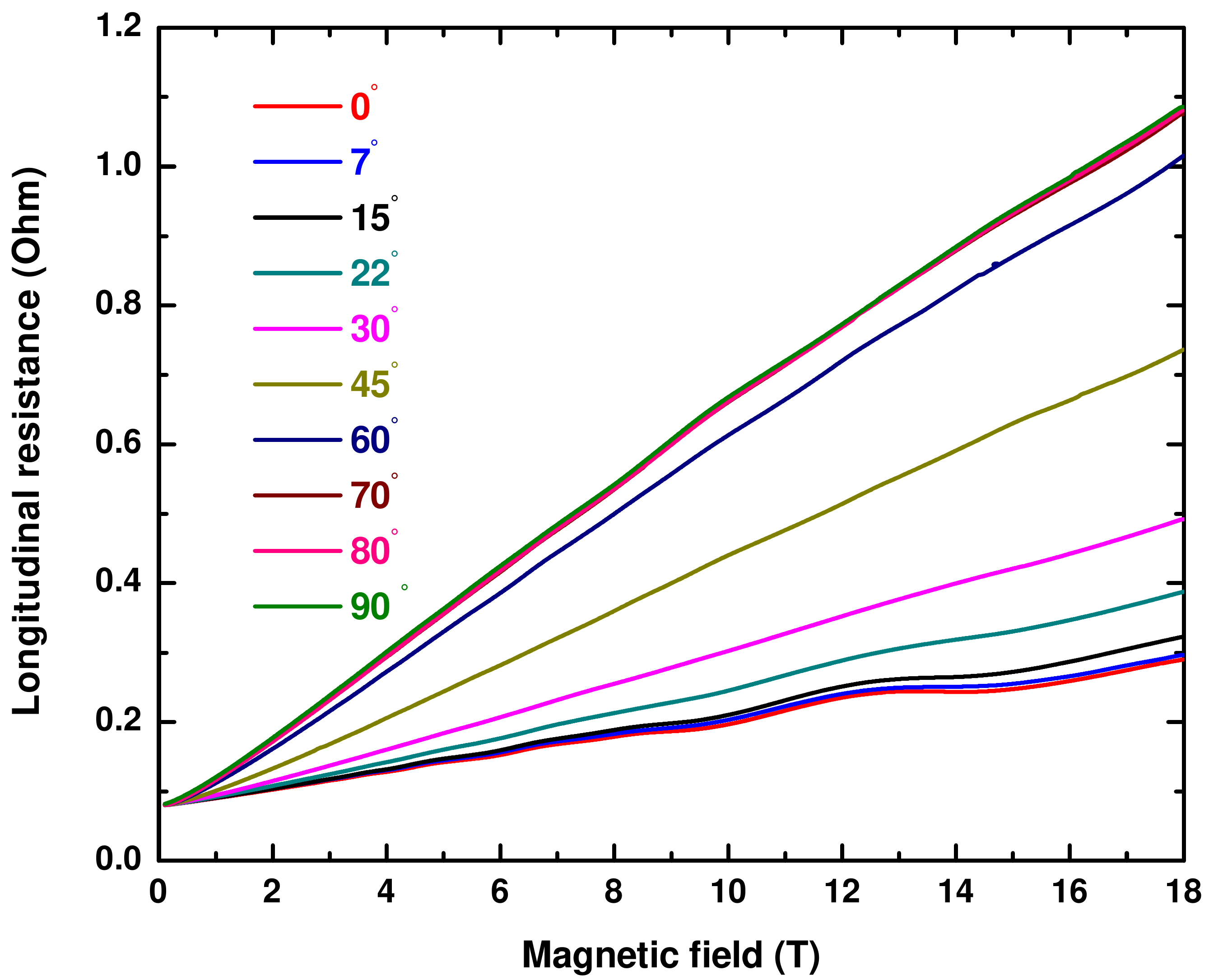}
\caption{\label{fig:angle} (Color online) Longitudinal resistance as a function of magnetic field for different tilt angle between normal to the surface and the magnetic field.}
\end{figure}

\begin{figure}[H]
\centering
\includegraphics[width=5 in,height=5 in,keepaspectratio]{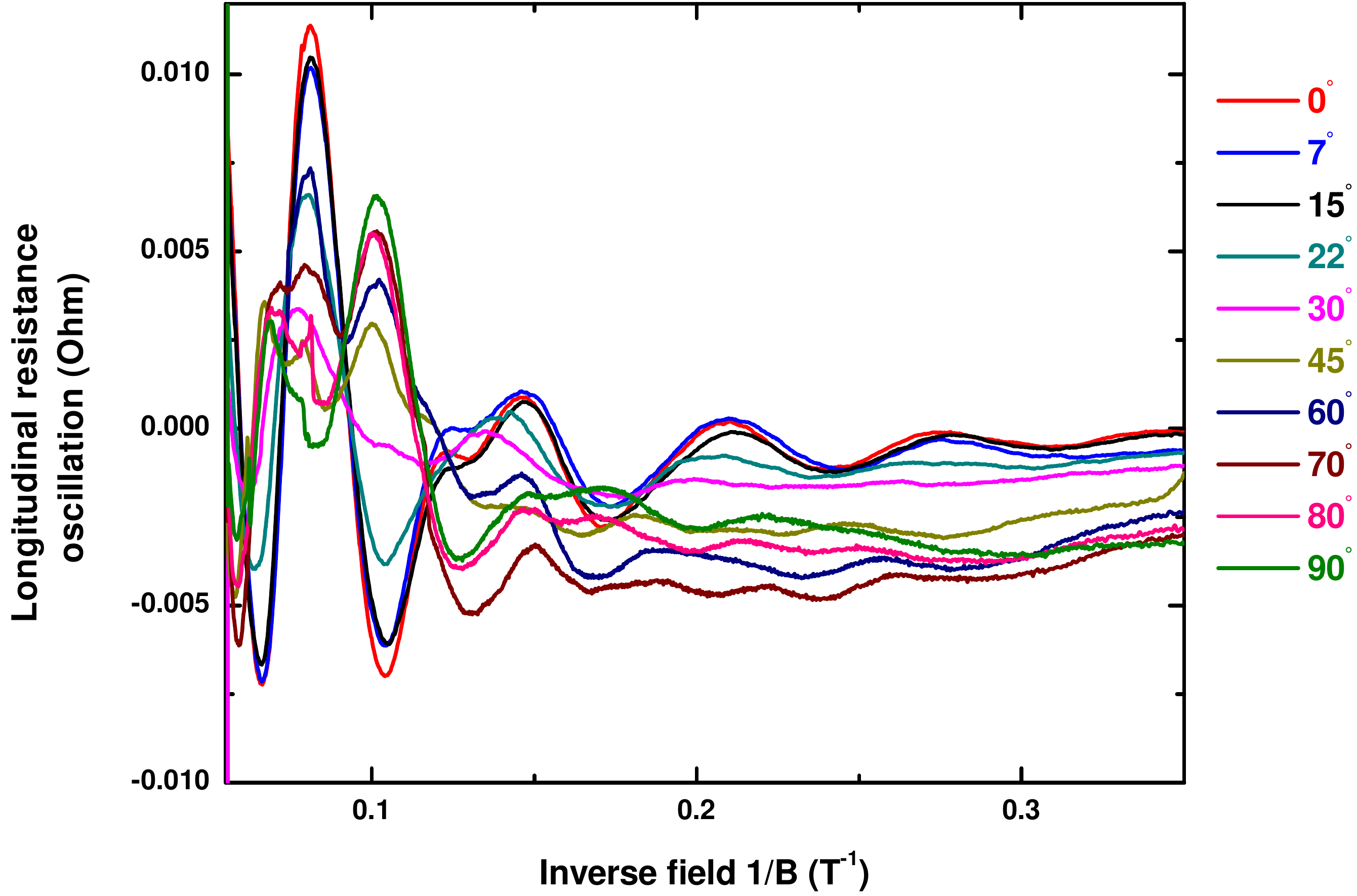}
\caption{\label{fig:allangle} (Color online) Longitudinal resistance oscillation as a function of magnetic field for different tilt angle between normal to the surface and the magnetic field.}
\end{figure}

\begin{figure}[H]
\centering
\includegraphics[width=4.5 in,height=4.5 in,keepaspectratio]{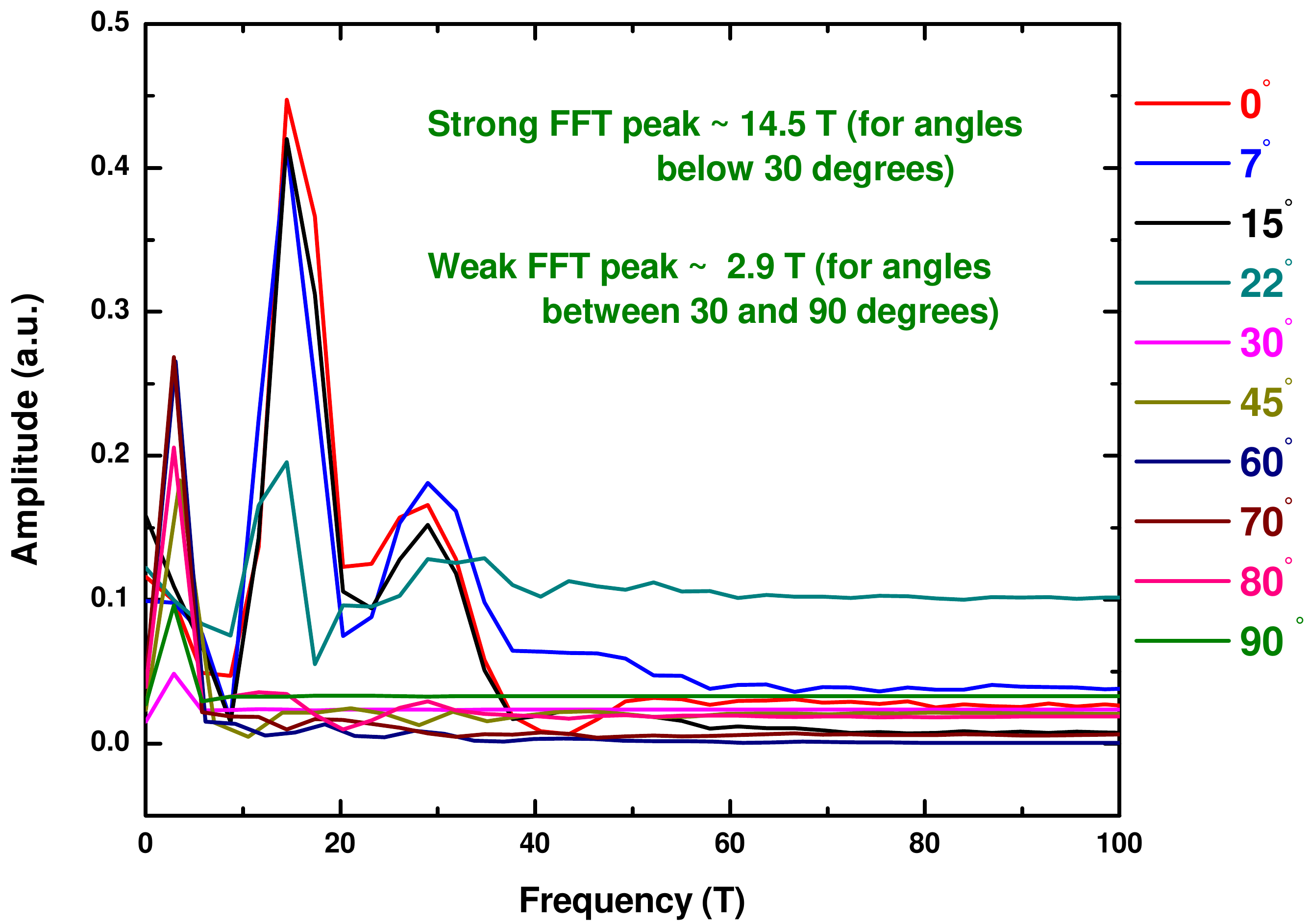}
\caption{\label{fig:FFTallangle} (Color online) FFT of longitudinal resistance oscillation for different tilt angles.}
\end{figure}

\subsection{Optical Study and Density Function Calculation}

Room temperature reflectance of PbSe between 35 and 35,000 cm${}^{-1}$ (4.3~meV--4.3~eV) is shown in Fig. 19(A). Towards the low end of far-infrared range, moderate reflectance of about 75\% is observed, decreasing quickly as frequency increases. The effect of an optically-active transverse  phonon may be inferred around 40--60~cm${}^{-1}$. There is a plasma minimum around 200 cm${}^{-1}$ and reflectance decreases to about 25\%. We used the Kramers-Kronig relations for the bulk reflectance $R(\omega)$ to estimate the real and imaginary parts of the dielectric function.\cite{Wooten}  Before calculating the Kramers-Kronig integral, the low frequency
reflectance data were extrapolated to zero using the reflectance-fit parameters discussed later in this section. Reflectance data above the highest measured frequency were extrapolated between 80,000 and $2\times10^{8}$ cm${}^{-1 }$ with the help of X-ray-optics scattering functions; from the scattering function for every atomic constituent in the chemical formula and the volume/molecule (or the density) one may calculate the optical properties in the X-ray region.\cite{Henke,Tanner} A power-law in $1/\omega$ was used to bridge the gap between the experimental data and the X-ray extrapolation. Finally, an $\omega^{-4}$ power law was used above $2\times10^{8}$~cm${}^{-1 }$.

\begin{figure}[H]
\centering
\includegraphics[width=5 in,height=5 in,keepaspectratio]{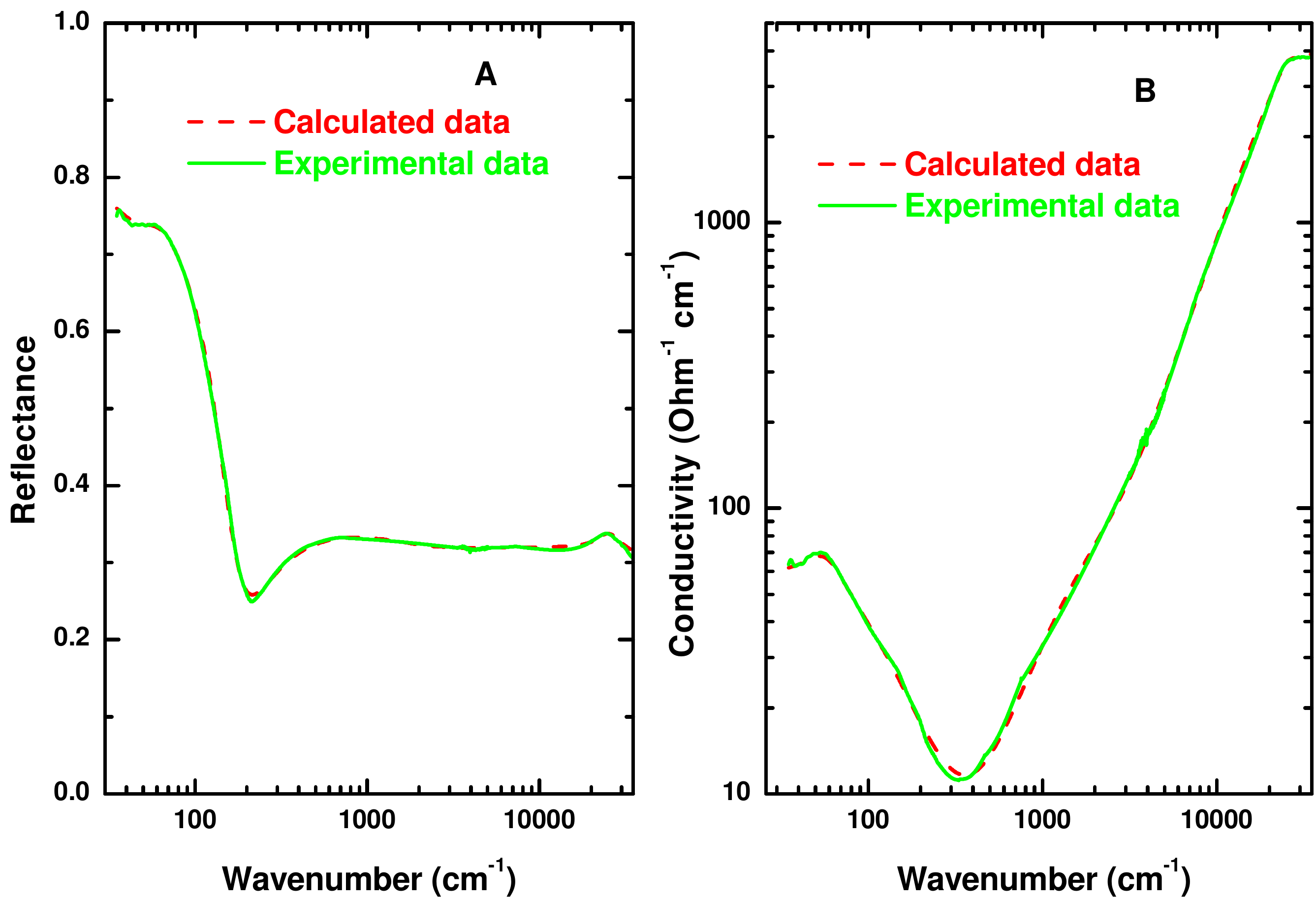}
\caption{\label{fig:optics} (Color online) A) Room temperature reflectance spectra and Drude-Lorentz model fit. B) A comparision between Kramers-Kronig derived and Drude-Lorentz parameter calculated optical conductivity.}
\end{figure}

The optical conductivity were derived from the measured reflectance and the Kramers-Kronig derived phase shift on reflection. The Figure 19(B) shows the Kramers-Kronig derived optical conductivity at 300~K. A small Drude component and phonon feature could be seen in the far infrared range whereas the higner frequency range consists of larger contribution to the conductivity from interband transitions, consistent with the reflectance spectra as well. Figure 19 also shows the Drude-Lorentz model fit to the reflectance and conductivity spectra. The Drude-Lorentz dielectric function is written as:

\begin{equation}
\varepsilon (\omega)=\varepsilon_{\infty}- \frac{\omega _{p}^{2} }{\omega^{2}+i\omega /\tau } +
\sum _{j=1}^3\frac{\omega_{pj}^{2} }{\omega_{j}^{2}-\omega^{2} -i\omega \gamma_{j} }
\end{equation}

Here the first term represents the core electron contribution (transitions above the measured range), the second term is free carrier contribution characterized by Drude plasma frequency $\omega _{p}$ and free carrier relaxation time $\tau $ and the third term is the sum of three Lorentzian oscillators representing phonons, and interband electronic contributions. The Lorentzian parameters are the $j$th oscillator plasma frequency $\omega _{pj}$, its central frequency $\omega _{j}$, and its linewidth $\gamma _{j}$. This dielectric function model is used in a least-squares fit to the reflectance. The Drude component characterizes the free carriers and their dynamics at zero frequency whereas the Lorentz contributions are used for the optically-active phonon in the far-infrared region along with interband transitions in the higher frequency region. A good agreement between KK derived conductivity and DL parameter calculated conductivity gives us confidence in the analysis procedure. The parameters are the ones used to fit the reflectance. These parameters are listed in table 4-1.

\begin{table}[h]
\caption { Drude-Lorentz parameters for PbSe at room temperature (300~K). }
\centering
\begin{tabular}{c c c c}
\hline
Modes assignment & Oscillator  & Central  & Linewidth \\
in fitting & strength $\omega _{p}$ & frequency $\omega _{j }$  & $\gamma _{ j}$   \\
routine  &  (cm${}^{-1}$) & (cm${}^{-1}$) &  (cm${}^{-1}$) \\[1ex]
\hline
Drude & 650 & 0 & 115  \\
Optical phonon & 200 & 55 & 35 \\
Interband 1 & 3100 & 2100 & 4500 \\
Interband 2 & 17,850 & 24,900 & 9800 \\ [1ex]
\hline
\end{tabular}
\end{table}

There is an optically active phonon mode around 55 cm${}^{-1}$ which is consistent with previously reported phonon frequency,\cite{Zhangyi,Kilian,TianZ} representing the 3-fold degenerate (cubic symmetry) transverse vibration of Pb atoms against Se atoms. Interband 1 mode represents the minimum absorption edge or the optical band gap of PbSe. The onset of this interband transition is around 2000 cm${}^{-1}$ or 0.25 eV which is in good agreement with previously reported band gap of PbSe.\cite{Delin,Otfried,Harman,Hummer}

Density functional theory (DFT) calculations were performed by Dr. Kamal Choudhary, Materials Science and Engineering UF, using the Vienna Ab-initio Simulation Package (VASP) and the projector augmented wave (PAW) method on a 2 atom primitive cell of PbSe.\cite{Kress,Kresse} Calculations were done using GGA-PBE\cite{Perdew} functional as well as hybrid functional HSE06.\cite{Heyd}, because PBE is generally prone to underestimate electronic band-gap of materials as shown in fig. 20.

\begin{figure}[H]
\centering
\includegraphics[width=5 in,height=5 in,keepaspectratio]{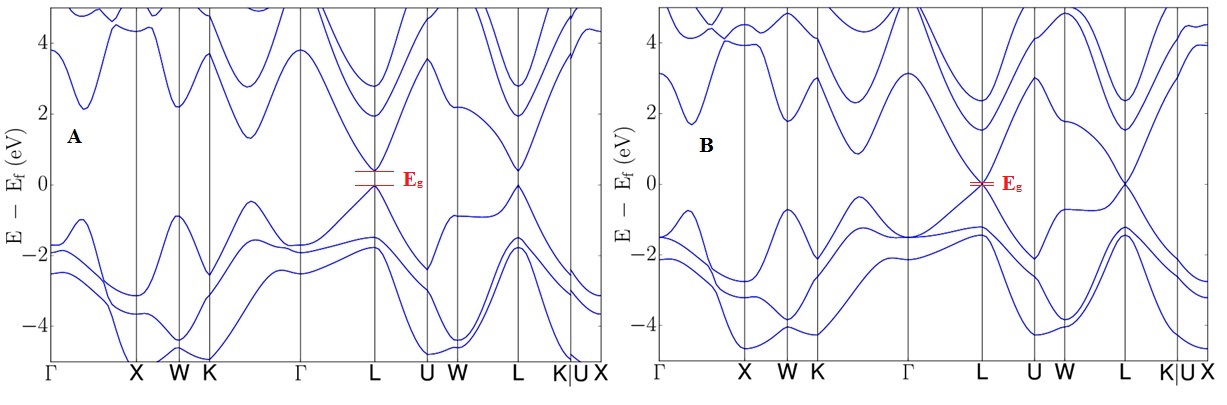}
\caption{\label{fig:bandpbse} (Color online) A) VASP calculated band structure of PbSe using hybrid functional HSE06. Predicted band gap is E$_{g}$ $\approx 0.26$ eV B) VASP calculated band structure of PbSe using GGA-PBE. Predicted band gap is E$_{g}$ $\approx 0.08$ eV. Figure courtesy by Dr. Kamal Choudhary.}
\end{figure}

Spin-orbit coupling was taken into account in both the cases. The conduction band minima and valence band maxima was found at L-point in the Brillouin zone as shown in band-structure. The valence band was dominated by Se, while the conduction band was dominated by Pb states as shown in element projected density of states in Fig. 21. The predicted direct band gap from hybrid functional was found to be around 0.28 eV while from GGA-PBE it was predicted to be around 0.08 eV. Obviously, VASP predicted band gap using hybrid functional appears in good agreement with experimental result.

\begin{figure}[H]
\centering
\includegraphics[width=5 in,height=5 in,keepaspectratio]{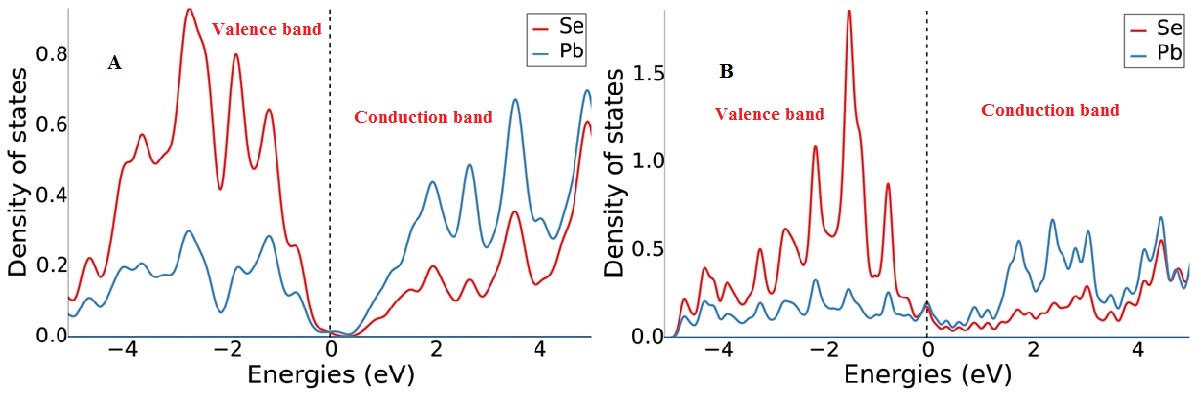}
\caption{\label{fig:dos} (Color online) A) VASP calculated density of states of PbSe using hybrid functional HSE06. Valence band is more Se-type whereas conduction band is more Pb-type. B) VASP calculated density of states of PbSe using GGA-PBE. Like hybrid functional, GGA also predicts valence band more Se-type whereas conduction band more Pb-type. Figure courtesy by Dr. Kamal Choudhary.}
\end{figure}

\section{CONCLUSIONS}

Hall measurements of PbSe single crystals disclose the p-type nature of the material and carrier density is around 1.6$\times10^{18}$~cm${}^{-3}$ in the temperature range of 0.8 K and 25 K. Low temperature and high magnetic field electrical properties measurement shows high positive magneto-resistance of at least 300\% below 25 K. At high magnetic field, longitudinal and Hall resistance shows oscillatory behavior also known as Shubnikov-de-Haas oscillations. The Lifshitz-Kosevich theory\cite{shoenberg} allows us to analyze the transport data. The frequency of oscillation of longitudinal and Hall resistance data is estimated to be around 14.5 T and oscillations are out of phase at high magnetic field. The effective mass of the charge carriers participating in the oscillation phenomena is estimated to be around 0.09$m_{e}$. Landau level fan diagram estimates the value of phase factor close to zero indicating that the charge carriers are non-Dirac fermions originating from bulk states. The free carrier scattering rate, mobility and conductivity is estimated to be around 26 cm$^{-1}$, 3990 cm${}^2$/Vs and 1020 $\Omega^{-1} cm^{-1}$. Dingle temeprature based on transpoprt scattering rate is estimated to be around 7 K. Angle dependent resistance measurement also shows oscillatory trend for high angles. Oscillation behavior and frequency of oscillation seems similar or in principle unchanged for angles below 30$^{\circ}$. Oscillation frequency is estimated to be around 14.5 T in this range. Oscillation pattern changes for angles greater than 30$^{\circ}$and oscillation frequency decreases to 2.9 T for angles between 30$^{\circ}$ and 90$^{\circ}$. Since oscillations appear for high angles indicate that Fermi surface is 3-Dimensional which is also predicted in previous study.\cite{Svane} Optical measurement and DFT calculation predicts that PbSe is a low-gap semiconductor with direct band gap at L-symmetry point in BZ of about 0.25 eV. An optically active transverse phonon mode is identified around 55 cm$^{-1}$. At 300~K, the screened Drude plasma minimum lies around 200 cm${}^{-1 }$. VASP calculated density of states predicts that the valence band in PbSe is mainly dominated by Se states, while the conduction band is dominated by Pb states.

\bibliography{PbSe}
\end{document}